\documentclass[aps,prb,twocolumn,showpacs,preprintnumbers,amsmath,amssymb]{revtex4}

\usepackage{dcolumn}
\usepackage{bm}
\usepackage[dvips]{graphicx}

\begin{document}

\preprint{APS/123-QED}

\title{Heterogeneous nucleation and adatom detachment at 1-D growth of Indium on Si(100)-2$\times$1}

\author{Jakub Javorsk\'y}
\author{Martin Setv\'in}
\author{Ivan O\v{s}t'\'adal}
 \email{ivan.ostadal@mff.cuni.cz}
\author{Pavel Sobot\'ik}
\affiliation{Department of Surface and Plasma Science, Charles University, V Hole\v sovi\v ck\'ach 2, 180 00 Prague 8, Czech Republic.}
\author{Miroslav Kotrla}
\affiliation{
Institute of Physics, Academy of Sciences of the Czech Republic, Na Slovance 2, 182 21 Prague 8, Czech Republic}

\date{\today}

\begin{abstract}
Growth of atomic indium chains - 1D islands - on the Si(100)-2$\times$1 surface was observed by scanning tunneling microscopy (STM) at room temperature and simulated by means of a kinetic Monte Carlo method. Density of indium islands and island size distribution were obtained for various deposition rates and coverage. STM observation of growth during deposition of indium provided information on growth kinetics and relaxation of grown layers. Important role of C-type defects at adsorption of metal atoms was observed. Measured growth characteristics were simulated using a microscopic model with anisotropic surface diffusion and forbidden zones along the metal chains. An analysis of experimental and simulation data shows that detachment of indium adatoms from the chains substantially influences a growth scenario and results in monotonously decreasing chain length distribution function at low coverage. Diffusion barriers determined from the simulations correspond to almost isotropic diffusion of indium adatoms on the surface. The results are discussed with respect to data reported in earlier papers for other metals.
\end{abstract}


\maketitle

\section {Introduction}

The Si(100)-2$\times$1 surface is composed of silicon atom pairs -- dimers -- arranged into rows. It represents a natural template for spontaneous growing linear structures of many materials, like group III-V metals. Group III metals (Al, Ga, In, Tl) are known to grow in one-dimensional (1D) atomic chains when deposited on the Si(100) surface \cite{1,2,3,4}. Technique of scanning tunneling microscopy (STM) enabled detailed study of metal layers with atomic resolution. An especially powerful tool for investigation of growth kinetics is the STM $in$-$vivo$ technique \cite{5,6} which allows direct monitoring of the layer as it grows. 1D metal chains grow perpendicularly to the underlying silicon dimer rows of the Si(100)-2$\times$1 surface. They are composed of metal dimers (oriented parallel to the silicon dimers) sitting in the ``trenches'' on the silicon surface \cite{7}. Growth of the metal chains has been explained by a surface polymerization reaction \cite{8}. Metal chain ends act as nucleation centers. Since the sites adjacent to a chain are energetically unfavorable for adsorption (no adsorption has been observed there) the chains grow only in length. The chains are always separated by a distance of at least 2a (a$=0.384$ nm, surface unit cell spacing). Thus, the surface is saturated by metal adatoms at a coverage of 0.5 ML (1~ML$= 6.78\times10^{14}$ cm$^{-2}$). Some differences exist between various group III metals. While Al and Ga chains are believed to be stable at room temperature \cite{8, 10}, indium atoms are known to detach from chains and re-attach to other chains \cite{9} and Tl chains were shown to be even more unstable \cite{4}.
Though a qualitative description of diffusion and growth processes for group III metals exists, the corresponding values of microscopic parameters are not known. The heights of diffusion barriers on Si(100) have not been yet  determined for In and the values reported by Albao $et~al.$ \cite{10} for Ga and by Brocks $et~al.$ \cite{8} for Al are very different. Similar discrepancy exists for estimation of Ga dimer pair-interaction energy -- Tokar and Dreyss\'e \cite{13} suggest $\approx$~0.2~eV while Takeuchi's ab-initio calculation gave $\approx$~0.8~eV \cite{16}. Recently Koc\'an $et~al.$ reported \cite{9} that detachment of indium atoms from chains is length dependent, so interactions other than nearest--neighbor (NN) might play a role.

The role of surface defects present on the Si(100)-2$\times$1 surface \cite{defects} at the metal adsorption and nucleation was reported and discussed \cite{9, commentKoc, albao_reply, Lesek}. Experimental results showed that influence of A- and B-type defects (one and two missing dimers, respectively) can be neglected but C-type defects are important. The C-type defects, which appear on STM images as a small bright protrusion next to a dark spot in filled states and as a larger bright spot in empty states, are reported to be very reactive and act as nucleation centers \cite{9,commentKoc}. The C-defects were independently interpreted by Hossain $et~al.$\cite{hossain} and Okano and
Oshiyama \cite{okano} as dissociated H$_{2}$O molecules, with the H and hydroxyl group bonded to neighboring silicon atoms of two adjacent surface dimers. The results both of experimental and theoretical study of In nucleation at the C-defects were reported in Ref.~\onlinecite{Lesek}. After adsorption of an In adatom at a C-defect (exclusively on the unoccupied side of the two adjacent Si dimers) the chain begins to grow in one direction only (see Fig.~1 in Ref.~\onlinecite{Lesek}). The chain termination at the defect is stable, the opposite ``free'' end is active as an adsorption site for adatoms.

Albao $et$~$al.$ \cite{10} reported growth characteristics for Ga on Si(100) at room temperature (RT). An unconventional monotonously decreasing scaled island (chain) size distribution function obtained for low coverage was explained by an irreversible growth model and kinetic Monte Carlo (KMC) simulations. The simulations resulted in the monotonous size distribution only if highly anisotropic surface diffusion of Ga adatoms was introduced otherwise a monomodal form of the distribution function was obtained. The presence of C-defects included later \cite{commentKoc,albao_reply} did not change the results significantly.
Similar growth characteristics at RT we reported for indium \cite{Ost_Kinetics_of_Growth}. The chain length distributions obtained for various coverages are monotonously decreasing and obey a scaling relation.  Most of In chains (60--90 \%) in observed layers were on at least one end terminated by a C-type defect. In the $in$-$vivo$ experiments the percentage was higher, 90--100 \% (due to low deposition rates of In during the $in$-$vivo$ measurements). Density of indium chains and average chain length depend on C-defect concentration. Another phenomenon (observed at higher coverage) is that indium atoms are able to migrate throughout sites adjacent to an indium chain even though no adsorption is observed (such an effect seems to be negligible at low coverage).

In this paper, we use STM data and KMC simulations for detail studying indium growth on the reconstructed surface Si(100)-2$\times$1. A growth model, which includes presence of C-defects and a process of atom detachment from indium chains, is used for studying a role of C-defects at metal chain nucleation, determination of diffusion barriers and investigation of relation between growth processes and a form of chain length distribution function.
Processes and parameters included in the growth model are discussed with respect to experimental data obtained by means of STM.

Performed experiments are characterized in Sec. II, consequently
experimental results are presented in Sec.~III. A simulation model is
described in Sec.~IV, results of simulations are compared with the
experimental data and discussed  in Sec.~V, formulas for calculation of
deviations between experimental and simulated growth characteristics
are given in the Appendix. Finally, Sec.~VI contains our conclusions.

\begin{figure}
\includegraphics[width=8cm]{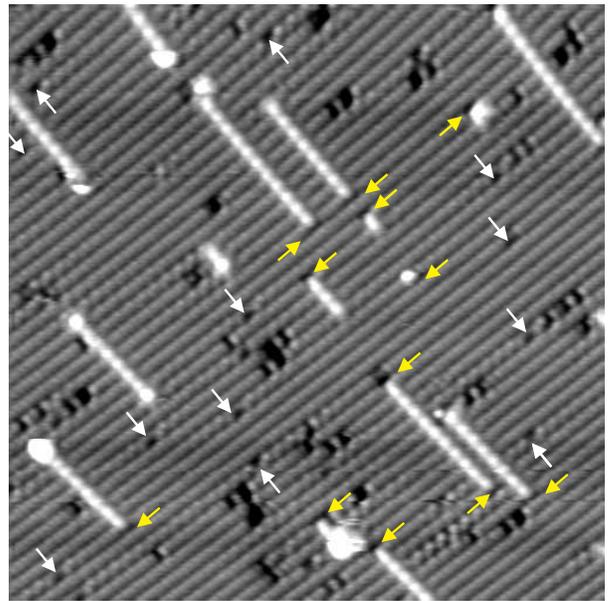}
\caption{Filled state STM image of indium 1D chains grown at room temperature on the surface Si(100)-2$\times$1. The bright spots at ends of some chains indicate termination by a single indium atom \cite{9,Lesek}; stable terminations at C-defects are marked by the arrows perpendicular to chains. Unoccupied C-defects are marked by the arrows parallel with chains -- the arrows are directed from the side of possible chain growing. Coverage 0.044~ML, $in$-$situ$ measurements 2 hours after deposition; image size 30$\times $30 nm$^{2}$. }
\label{STMimage}
\end{figure}

\section {Experimental}
All experiments were carried out at room temperature in a non-commercial UHV STM system with base pressure $< 3\times10^{-9}$ Pa. Si(100) samples were cut from an n-type, Sb doped silicon wafer with resistivity $\le$ 0.014 $\Omega$cm. To obtain the 2$\times$1 reconstruction, samples were flashed several times for $\approx$ 20 $s$ to $\approx$ 1200$^\circ$ C. Indium was deposited from tungsten wire evaporators either before or during imaging the surface by means of STM. In the latter case (experiment {\it in-vivo}) a miniature evaporator was in a distance of 4 cm from the sample and a beam of In atoms was determined by means of two apertures with a diameter of 1~mm. Incidental angle of the beam was 30$^\circ$. The apex shape of a tungsten electrochemically etched tip enabled deposition ``under'' the STM tip onto the scanned area. The thermal drift during deposition was compensated by the STM control unit. The $in$-$vivo$ measurements provided continuous STM imaging of the investigated surface area before and during the deposition at a rate of 1 image/min. At standard $in$-$situ$ measurements the indium layers were observed 0.5--4 hours after deposition. We used a tip voltage of +2 V and tunneling current 0.3 nA, the values at which tip influence on the detachment of indium adatoms from the chains is minimized as we proved earlier \cite{9}.

\section{Experimental results}

An example of indium chains grown on the Si(100) surface at low
coverage is shown in Fig.~\ref{STMimage}. Length of the chains can be
easily determined with atomic precision from filled states images where
the ``free'' terminations of chains containing an odd number of atoms
appear much brighter than in case of even number \cite{9,commentKoc}.
Chain terminations at C-defects and unoccupied C-defect are marked by
arrows. Concentration of the C-defects increases during deposition at
$in$-$vivo$ experiments (probably due to thermal desorption of residual
water molecules from heated parts near the evaporation source in a
relatively small distance from the sample). The increase was found
linear and corresponds to a deposition rate of $4\times10^{-6}$ ML/s.
The deposition rate of C-defects at $in$-$situ$ experiments was
negligible (due to a large distance of the evaporation source and better
screening of the sample) and the initial concentration of the C-defects
can be considered as unchanged.

Since the atomic structure of In chains is well known \cite{7, 9, Lesek} we focused on statistical characteristics of the chains. Obtained data are summarized in the Table \ref{tab:table1}. The data were acquired from images of the size of 30$\times$30 nm$^2$ or 40$\times$40 nm$^2$. The image areas are large enough (30 nm corresponds to a $\sim$ 100-atoms long chain) and the resolution is sufficient to discern the number of atoms in the chain.
Data were collected only from terraces much wider than an average chain length to exclude the influence of step edges on the chain growth. Statistical characteristics were evaluated for both types of growth experiments ($in$-$vivo$ and $in$-$situ$).

\begin{table*}
\caption{\label{tab:table1}Measured statistical characteristics of In chains for various coverages and deposition rates.}
	\begin{ruledtabular}
	\centering
		\begin{tabular}{p{0.3in}p{0.3in}p{0.5in}p{0.7in}p{0.7in}p{0.3in}p{0.5in}p{0.5in}p{0.7in}}
Coverage&	Average chain length&	Average length (chains with free ends)&	Average length (chains terminated on C-defects)	& Percentage of chains terminated on C-defects&	Total number of chains&	Deposition rate (ML/s)&	C-defect concentration (ML$^{-1}$)&	Percentage of occupied C-defects\\
\hline
0.01&	2.78&	 2.87&	2.77&	 0.90&	303		&0.03    &0.014	 &0.14\\
0.04&	4.19&	 3.48&	4.63&	 0.61&	1098	&0.0035	 &0.008	&0.74\\
0.05&	4.92&	 6.00&	4.57&	 0.75&	154	  &0.01	   &0.013	&0.58\\
0.08&	6.64&	 5.14&	6.95&	 0.83&	548	  &0.002  &0.011	&0.93\\
0.09&	8.20&	 6.71&	8.69&	 0.75&	69	  &0.0045 &0.013	&0.68\\
0.15&	18.29& 19.29&	17.60&	0.59&	207	&0.003	&0.005	&0.94\\
		\end{tabular}
	\end{ruledtabular}	
\end{table*}

\begin{figure}
\includegraphics[width=8cm]{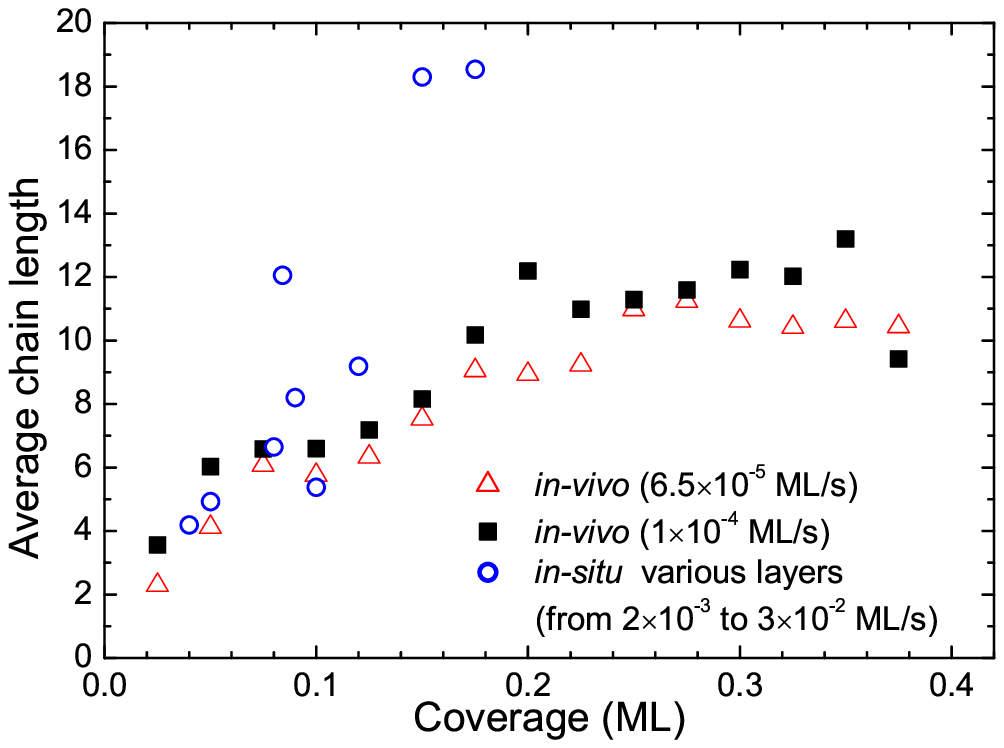}
\caption{Dependence of average chain length on coverage obtained from STM experiments. The average length of In chains increases with coverage and saturates at a coverage of 0.25 ML for the $in$-$vivo$ data. The dependence increases faster for the layers prepared at $in$-$situ$ experiments due to smaller and constant concentration of C-defects. The effective deposition rate for C-defects was 4$\times10^{-6}$ ML/s during the $in$-$vivo$ measurements. }
\label{expdata}
\end{figure}

Length of the grown indium chains evolves in time due to attachment/detachment of single atoms to/from the ``free'' ends. Fig.~ \ref{expdata} contains dependence of an average chain length on coverage obtained from
two $in$-$vivo$ measurements with deposition rates 6$\times10^{-5}$ and 1$\times10^{-4}$ ML/s and a number of $in$-$situ$ measurements for various deposition rates (from 2$\times10^{-3}$ to 3$\times10^{-2}$ ML/s) and coverages (0.01--0.15 ML).
The average length of chains is smaller in case of $in$-$vivo$ experiments because of higher concentration of C-defects, and from the coverage 0.25~ML (which corresponds to occupation of a half of all possible adsorption sites on the surface) almost does not increase.

\begin{figure}
\includegraphics[width=8cm]{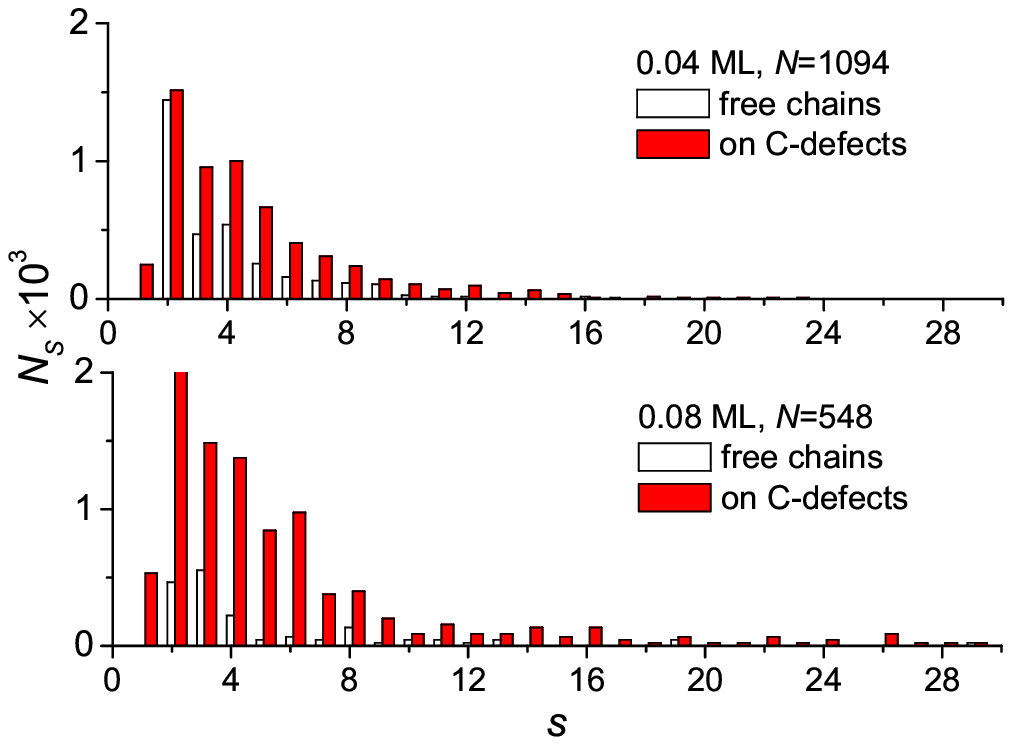}
\caption{Histograms of chain length distribution obtained for two indium layers with coverage of 0.04 and 0.08~ML deposited at a deposition rate of $\approx$0.004 and $\approx$0.002~ML/s respectively ($in$-$situ$ experiments). For the comparison histograms contain single indium atoms trapped on C-defects. $N$ represents a number of investigated chains.}
\label{histograms}
\end{figure}

Fig.~\ref{histograms} shows histograms of chain length distributions for two different coverages obtained from $in$-$situ$ experiments. The data were averaged over STM images taken after the deposition from various surface areas. The chains nucleated on C-defects are distinguished from the ``free'' chains (without termination on a C-defect). The histograms contain also single In atoms adsorbed and trapped on C-defects. They represent stable objects with a role of nucleation centers. Their presence in the histograms allows better understanding of the growth but we consider only chains with length $s \geq 2$ as parts of the investigated ``island population''. The histograms are decreasing for $s>1$ and the same tendency was observed for the other prepared layers with low coverages $\leq 0.15$~ML. The monotonously decreasing chain length distributions obtained for growth of indium at RT represent a remarkable quality. It can be seen that histograms contain some features related to details of the chain growth -- a small excess of chains containing even number of atoms which corresponds to higher stability of such chains experimentally observed \cite{9} and calculated \cite{Lesek} before.

\begin{figure}
\includegraphics[width=8cm]{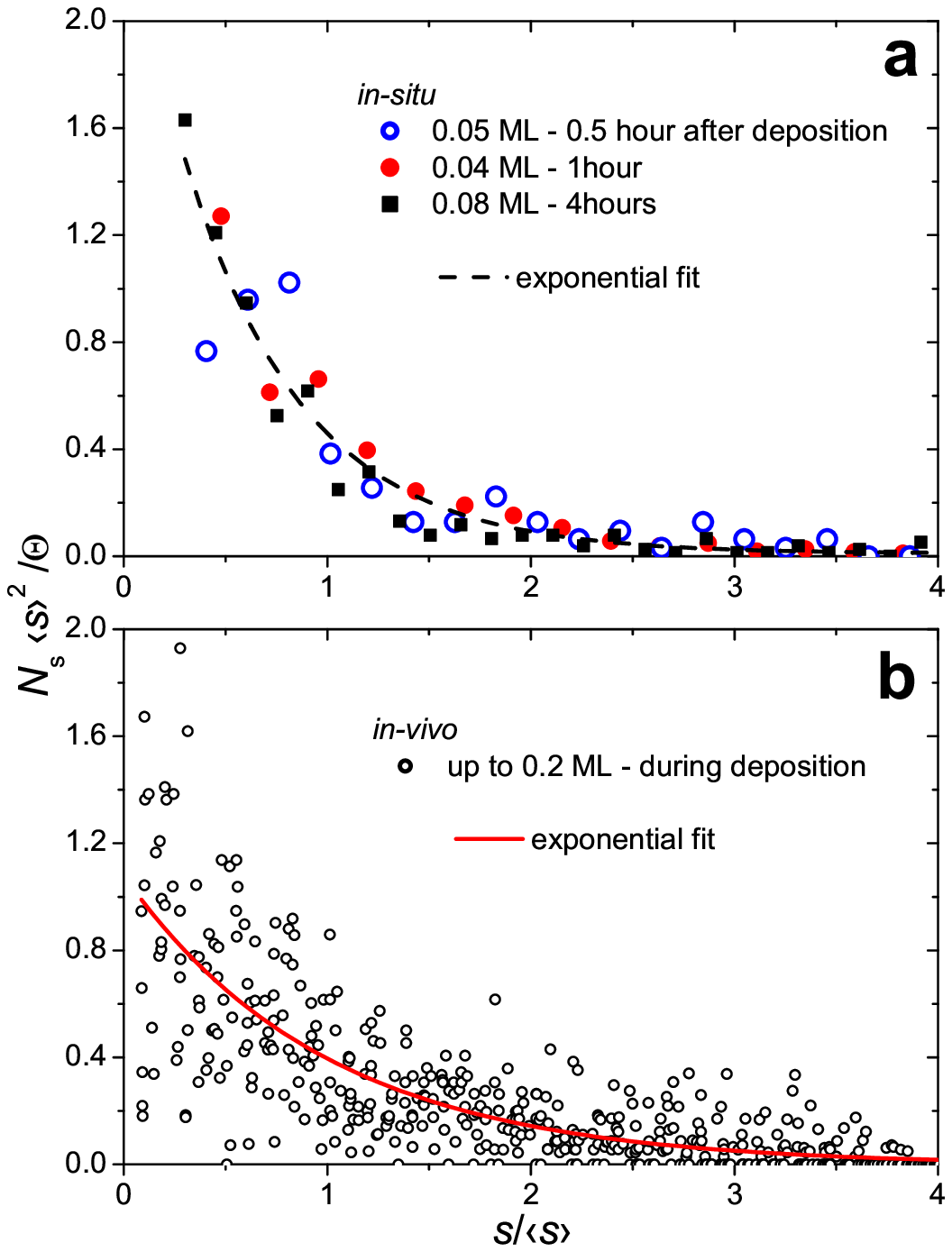}
\caption{Scaled island size distribution functions obtained from experimental data. (a) Results from three $in$-$situ$ measurements performed at various moments after deposition. (b) Data obtained during deposition ($in$-$vivo$) -- composed from a set of 15 histograms covering equally the whole deposition. The data were fitted by exponential functions.}
\label{expscaledfunction}
\end{figure}

The chain length distributions exhibit scaling \cite{14,15}: the function
$N_{s} \langle s \rangle ^{2}/\Theta$ scales with $s/\langle s \rangle$,
where $N_{s}$ is density per site of chains of the length $s$, $\langle s
\rangle$ -- average chain length and $\Theta$ -- coverage. The upper
panel (a) of Fig.~\ref{expscaledfunction} shows the scaled distribution
functions corresponding to various delay between the end of deposition
and STM measurements at the $in$-$situ$ experiments. Only chains
with $s\geq 2$ were included. The all data were fitted by an exponential
function. Due to statistical fluctuations in the data it is difficult to
distinguish reliably an effect of postdeposition relaxation. The postdeposition relaxation can be expected
because of the process of detachment, which introduces a feature of
``reversibility'' into a growth mechanism. We will discuss the
``reversibility'' and postdeposition relaxation later in Section V.

The scaled distributions obtained from the images recorded at {\it in-vivo} measurements are in the bottom diagram (b) in Fig.~\ref{expscaledfunction}. The data suffer from limited size of the investigated surface area and relatively small number of observed metal chains. Values for a particular distribution corresponding to a chosen moment (coverage) of the growth were averaged from a set of 3 subsequent STM images around that moment. The whole recorded sequence of images taken during the growth up to $0.2$~ML at a deposition rate of $0.0001$~ML/s (estimated from the images) is represented "equidistantly" by the 15 distributions. The rather scattered data exhibit monotonous character and can be approximated by an exponential function.

\begin{figure}
\includegraphics[width=8cm]{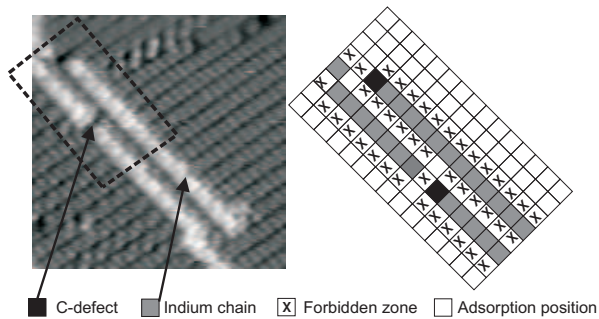}
\caption{Left: Filled state image of indium chains. The two neighboring chains are separated by a distance of one lattice unit. Dark spots at the ends of chains are C-defects. Right: Alignment of the model to Si(100) substrate. The black square represents the ``reactive site'' at the C-defect position, grey squares -- adsorption sites occupied by indium adatoms, empty squares -- unoccupied adsorption sites, crossed squares -- ``forbidden zones''-- energetically unfavorable adsorption sites.}
\label{model}
\end{figure}

\section{Simulation model}

We developed a physical atomistic model for submonolayer growth of indium on Si(100). Similarly as in the model used by Albao $et~al$. to describe growth of gallium layers at low coverage \cite{10}, we consider anisotropic diffusion of the metal adatoms on the $2 \times 1$ reconstructed surface. However, we took into account experimental results obtained for In and included new features: presence of C-defects acting as nucleation centers on the surface and detachment of single atoms from chains. The mechanism of detachment introduces into the growth model a possibility of ``reversible'' behavior (``reversible'' does not mean here ``equilibrium'' -- due to existing flux of deposited atoms). Approaching an equilibrium state depends on growth conditions and it competes with kinetics which controls growth entirely in case of an irreversible model used for gallium.

The model for In deals with four types of objects on the surface (see Fig.~\ref{model}):

1) Free In adatoms -- indium atoms perform thermally activated hopping on the surface represented by a square lattice (see Fig.~\ref{model}). The hopping is anisotropic (parallel and perpendicular to indium chains) described by rates
 $h_{\|,\bot}=\nu \exp[-E_{\|,\bot}/(k_{B}T)]$, where the attempt frequency was set as $\nu=10^{13}\;s^{-1}$, the activation energies (diffusion barriers) $E_{\|}$, $E_{\bot}$ are simulation parameters, $k_B$-- the Boltzmann constant and $T$-- temperature.

2) Forbidden zones -- surface sites not accessible for hopping adatoms. The forbidden zones were introduced -- similarly as in Ref. \onlinecite{10} -- to simulate the 1D growth, minimum separation observed between two adjacent chains and the fact that no chain nucleates at a ``non-reactive'' side of a C-defect.

3) C-type defects -- dissociated H$_2$O molecules -- rest on top of
dimer rows. STM observations show that one side of the C-defect acts
as a nucleation center while no adatom adsorption has been observed
on the opposite "dark" side (on filled state STM image) of the defect.
Thus, C-type defects are represented by a single forbidden site and an
adjacent ``reactive'' site.

4) Indium chains -- orientated perpendicularly to the Si dimer rows -- are composed of two or more In atoms. Similarly as in the model used by Albao $et~al$.\cite{10} arrangement of atoms in chains into dimers is not taken into account.

There are three main processes in our model:

A) Deposition -- Adatoms and C-defects are deposited randomly. If a defect or an adatom is deposited on an already occupied position or (in case of In atoms) into
a forbidden zone, the nearest free position is looked up in the direction of dimer rows. Deposition flux and time were set the same as in the particular experiments (indium flux $~10^{-2}-10^{-5}$ ML/s, C-defect flux $4\times10^{-6}$ ML/s -- for ``$in~vivo$'' only) and corresponding simulations were performed for all STM experiments. Orientation of a deposited C-defect (reactive site) was chosen randomly. As the C-defects change their state only very rarely \cite{Lesek,cdef} the orientation is fixed during the whole simulation (it differs from Ref.~\onlinecite{albao_reply}, where the orientation was determined each time when an adatom tried to hop next to the defect).

B) Surface migration (hopping) -- Single indium atoms deposited on surface perform random hops among adsorption sites. Jumps to forbidden zones and on top of other atoms are prohibited
An atom is trapped when hops onto a C-defect reactive site. If an atom hops on a site next to another atom in the direction parallel to In chains (perpendicular to the dimer rows), a new indium chain nucleates or an existing one grows. The hopping rates $h_{\|,\bot}$ in directions parallel or perpendicular to the indium chains are given by the simulation parameters $E_{\|} $ and $E_{\bot}$ (diffusion barriers parallel and perpendicular to indium chains).

C) Detachment -- An indium atom can detach from a chain end, not
terminated by a C-defect, by jumping off either in perpendicular or
parallel direction to the chain. According to our best knowledge there
are neither experimental nor theoretical data available to characterize
the detachment direction in the studied (or similar) system on the atomic
level. The process of the detachment is thermally activated and can be
described by two parameters: $E_{det\|} $ and $E_{det\bot} $ (energy
barriers for detachment in the parallel and perpendicular direction).  Our
previous measurements \cite{9} and theoretical calculations \cite{Lesek}
show that the energy for detachment depends on length and termination
of an In chain by a single atom or dimer. The detachment from a chain
containing an odd number of atoms is easier than in case of even number.
For the simplicity the model contains only one parameter -- the
value of a mean energy barrier for detachment, $0.82$~eV, derived from 
experimental data reported in Ref.~\onlinecite{9}.
As we know only the total rate of detachment
without any details, simulations were performed for detachment
either in the parallel or perpendicular direction. Two-atomic chains
(dimers) on C-defects are considered as stable objects which cannot
decay (both experiment and theoretical calculations \cite{Lesek}
confirmed their high stability).

The simulation proceeds as follows: C-defects are randomly deposited on the surface with initial concentration. Then indium atoms are randomly deposited (together with additional C-defects according to a simulated experiment). For details of an employed method of KMC simulations see Ref.~\onlinecite{Kotrla_KMC}. After the deposition stops the layer is allowed to relax
for the same time as in the corresponding experiment. The Si(100)-2$\times$1 surface is represented by a square matrix of adsorption sites, each capable of holding a single indium atom. We used a matrix size between 100$\times$100 and 500$\times$500 lattice units, each simulation was repeated at least 9 times and
the obtained data were averaged. The values of the matrix size were chosen so as the mean average error of simulated data was 2--3$\times$ smaller than that of measured data (size of statistical arrays varied for different values of coverage). The boundary conditions were periodic.

\section{Results of simulations and discussion}

{\it Diffusion barriers}.   The $in$-$situ$ and $in$-$vivo$ experiments were simulated for various combinations of diffusion
barriers $E_{\|} $, $E_{\bot}$.
Comparison of experimental and simulated growth characteristics (average chain length, dependence of average chain
length on coverage and scaled chain length distribution) was used for estimation of a combination
of barriers
which provides the best agreement -- the lowest deviation calculated as presented in the Appendix.

\begin{figure*}
\includegraphics[width=16cm]{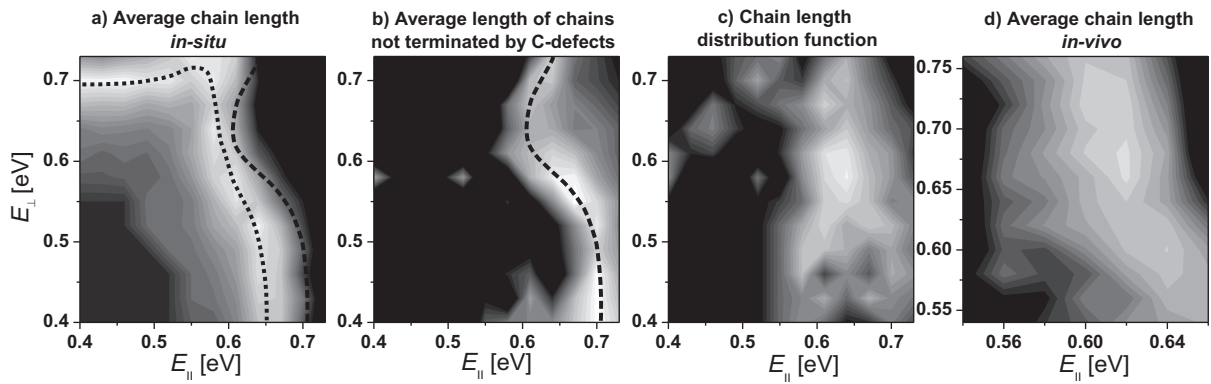}
\caption{Maps showing deviation between the measured and simulated characteristics of indium chains as a function of diffusion barriers parallel and perpendicular to In chains. White areas indicate the best agreement, black areas the worst. Dotted   line connects local minima in panel a), dashed line in panel b). Dashed line from panel b) is also shown in panel a) for comparison.  Energy barrier for atom detachment was 0.82~eV. Only detachment parallel to indium chains was allowed in these simulations.}
\label{mapy}
\end{figure*}

Figure~\ref{mapy} contains diagrams with the dependence of the
calculated deviations on the simulation parameters $E_{\|} $ and
$E_{\bot}$ for the chosen growth characteristics. The grayscale
represents accuracy of the match for a given combination of parameters
with white for the most precise match. The plots a) and b) demonstrate
fitting the $in$-$situ$ measurement for a layer with coverage 0.08~ML. In the plot~a) the
dotted black line shows the combinations of activation energies which
provide the best match between the measured and simulated average
length of indium chains, the dashed line in the plot~b) shows the best
agreement for the average length only of those chains which are not
terminated by C-defects at any of ends. The relatively small
``sub-population'' of these chains behaves in a different way so the two
plots together were used to determine the best combination of energies:
$E_{\|} $= 0.62$\pm$0.03 eV and $E_{\bot}$ = 0.61$\pm$0.07 eV. The
values are consistent (within the errors) with the optimum combination of
energies  obtained from fitting average chain length dependence on
coverage from $in$-$vivo$ measurements -- see Fig.~\ref{mapy}d).

\begin{figure}
\includegraphics[width=7cm]{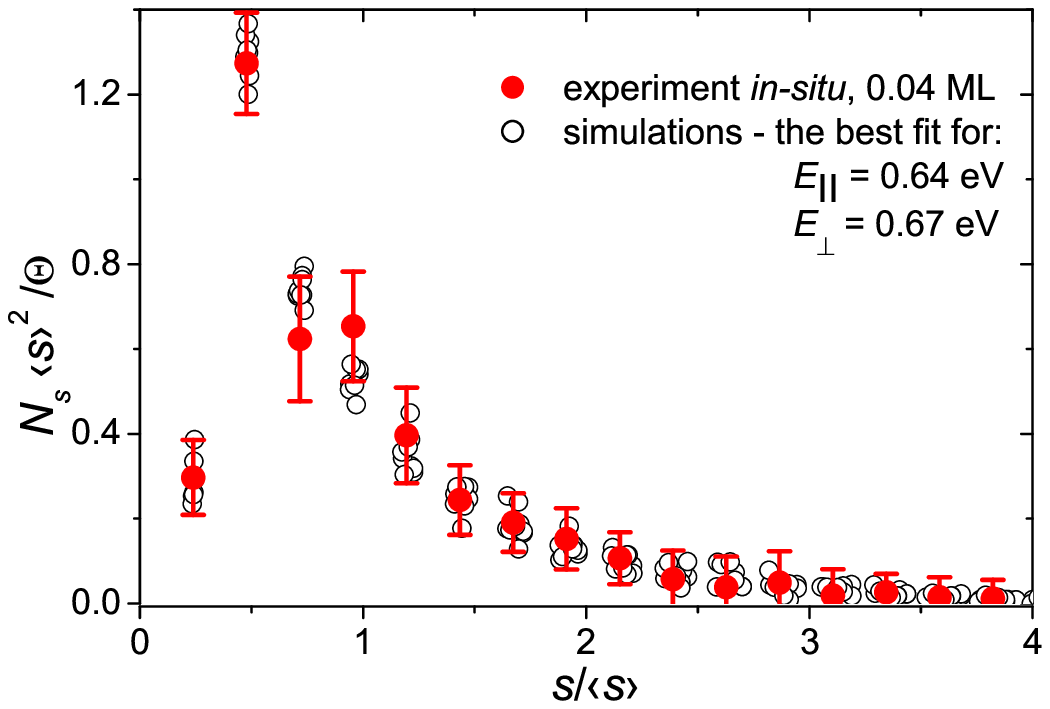}
\caption{Experimental and simulated scaled chain length distribution functions for a relaxed layer at coverage 0.04 ML ($in$-$ situ$ experiment). Results of various simulation experiments for the same parameters $E_{\|}$ and $E_{\bot}$ are plotted. Data representing single indium atoms trapped on C-defects are plotted as well (on the very left side).}
\label{scaled}
\end{figure}

Further we simulated experimentally obtained scaled chain length
distributions. Fitting was performed for the two $in$-$situ$
measurements which provided the largest data sets (0.04 and 0.08 ML
coverage) and for the $in$-$vivo$ measurements.
As an example of analyzed data we show in Fig.~\ref{scaled}
simulations of experimental data for a distribution function. 
The simulations were performed for the energies $E_{\|} = 0.64$~ eV
and $E_{\bot} = 0.67$ eV and correspond to a layer with coverage 0.04~ML (see Table~\ref{tab:table2}).   
The plot, in addition to points representing
the whole chain population for $s\geq 2$, contains the points
corresponding to single indium atoms trapped on C-defects. The
deviation used for the final fitting was a sum of $D_{iii}$ values
calculated using the relation (\ref{scalederror}) for the three chosen
measurements (see Fig.~\ref{mapy}c). Comparison of experimental and
simulated data resulted also in almost isotropic diffusion of In on Si(100)
with activation energies very close to the values obtained by the fitting
represented by the diagrams a), b) and d).

In addition, we investigated how the direction of detachment of
atoms from chains affects values of the estimated diffusion barriers. We found
that the direction of detachment does not affect results significantly
when diffusion is nearly isotropic, it plays an important role only in case
of very anisotropic diffusion.
Table \ref{tab:table2} shows the energy barriers obtained from simulations of different experiments, fitted using
both the parallel and perpendicular direction of detachment. Any combination of the energies outside the range given by
included errors results in a double deviation (compared to the best fit) between experimental and simulated data.
The activation energies obtained for the parallel detachment are slightly lower than for perpendicular one.
\begin{table}
\caption{\label{tab:table2}Activation energies for In adatom diffusion parallel and perpendicular to In chains for two directions of detachment. Values were obtained by comparing data from two {\it in-situ} and one {\it in-vivo} experiments and simulations for two models considering different atom detachment from chains.}
	\begin{ruledtabular}
		\begin{tabular}{lccccc}
			 &\multicolumn{2}{c}{Parallel detachment}& \multicolumn{2}{c}{Perpendicular detachment}\\
			 \hline
			 &E$_{\|}$ [eV]&E$_{\bot}$ [eV]&E$_{\|}$ [eV]&E$_{\bot}$ [eV]\\
			 0.04 ML &0.64 $\pm$ 0.03 &0.62 $\pm$ 0.07 &0.64 $\pm$ 0.03 &0.67 $\pm$ 0.07\\
			 0.08 ML &0.62 $\pm$ 0.03 &0.61 $\pm$ 0.07 &0.61 $\pm$ 0.03 &0.64 $\pm$ 0.07\\
			 in-vivo &0.60 $\pm$ 0.10 &0.65 $\pm$ 0.05 &0.65 $\pm$ 0.10 &0.65 $\pm$ 0.05\\
		\end{tabular}
	\end{ruledtabular}	
\end{table}

The diffusion barriers
for indium
determined from KMC simulations (see Table \ref{tab:table2}) correspond to almost isotropic diffusion.
It is in contrast
to
Albao's $et~al$. anisotropic results (E$_{\|}=0.40$~eV; E$_{\bot}=0.81$~eV). The vales obtained theoretically
for Al by Brock and Kelly \cite{8} are anisotropic as well (E$_{\|}=0.1$~eV; E$_{\bot}=0.3$~eV) but much lower. \\

{\it Scaled chain length distribution function.} It follows from the nucleation theory \cite{14,evba} that during the submonolayer
growth the density (per site) $N_s$ of islands composed of $s$ atoms fulfils the scaling form:

\begin{equation}
	N_s \approx \Theta\frac{f(s/\langle s \rangle)}{\langle s \rangle^2},
	\label{scaledfunction}
\end{equation}

where $\Theta$ is coverage and  $\Theta/\langle s\rangle$ represents
the mean island size density. The function $f(x)$, $x=s/\langle s\rangle$
is the scaled island size distribution function. The  relation
(\ref{scaledfunction}) was confirmed by simulations using different
models of irreversible 2D aggregation and from STM experiments. In
most cases the shape of the scaled distribution function is monomodal --
with a peak for $x=1$. 
A monomodal function was observed both for homogeneous and
heterogeneous nucleation. The scaling for the passage from irreversible
to reversible aggregation was examined theoretically in
Ref.~\onlinecite{15}. The unconventional shape of the scaled distribution
function -- monotonously decreasing -- observed for 1D growth of Ga on
Si(100) by Albao $et~al$. (and explained by means of KMC simulations
using strongly anisotropic surface diffusion) \cite{10} was theoretically
investigated by Tokar and Dreyss\'e \cite{13}. They showed that for
equilibrium growth and a model with atomic interactions restricted to
only nearest neighbors, the scaled distribution function is exponentially
decreasing. Here we obtained for 1D submonolayer growth of
In similar monotonously decreasing chain length distribution functions --
Fig.~\ref{expscaledfunction} and Fig.~\ref{scaled}.

If we consider homogeneous nucleation with the detachment only and the C-defects are excluded from our model
the simulated distribution function remains monotonously decreasing. On the other side excluding the process of
detachment (irreversible model) results in a conventional monomodal distribution function independently on presence
of the C-defects in the model. For the irreversible model the monotonous distribution can be simulated only when strong
diffusion anisotropy is introduced -- similarly as reported Albao $et~al$.\cite{10}

We conclude that the monotonous form of the chain length distribution function obtained
for indium layers with low coverage ($\leq0.15$~ML) 
at RT and used deposition rates can be explained by the process of atom detachment from indium chains during the
growth.\\

\begin{figure}
\includegraphics[width=8cm]{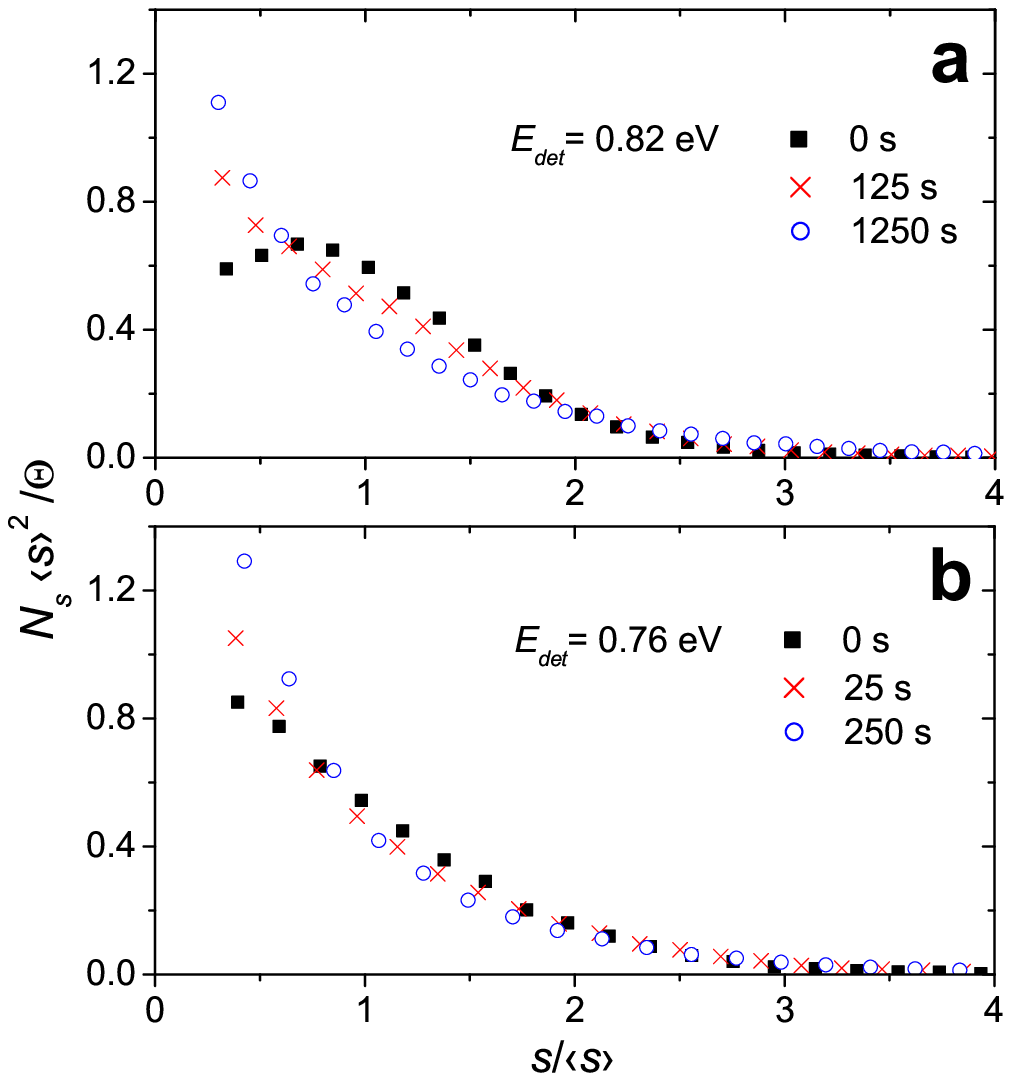}
\caption{Post-deposition relaxation of a simulated scaled distribution functions for two detachment barriers -- (a) 0.82 eV and (b) 0.76 eV. The transition to the monotically decreasing function corresponding to an equilibrium state is visible. The simulation corresponds to a layer with coverage 0.08~ML deposited at rate 0.002~ML/s.}
\label{relax}
\end{figure}

{\it Postdeposition relaxation}.  Further we used our reversible growth
model for exploring postdeposition relaxation indicated by experimental
data obtained from STM measurements at various time after deposition
(see Fig.~\ref{expscaledfunction}). We simulated the growth using the
diffusion barriers determined and deposition rate 0.002~ML/s for two
different values of the detachment barrier. The time evolution of the
distribution functions obtained for various time after deposition is in
Fig.~\ref{relax}a,~b. If the detachment rate is small enough with respect
to deposition rate
(a high energy barrier for detachment)
the distribution function is monomodal just after the
deposition and relaxes into a monotonous one -- Fig.~\ref{relax}a. If the
detachment rate increases the distribution function changes from the
monomodal to monotonous form even during the growth -- see
Fig.~\ref{relax}b. The simulation shows that the observed system
reaches an equilibrium state after $\sim$~6 hours (estimation for the used experimentally determined barrier 
$E_{det}= 0.82$~eV), though the most
dramatic change occurs during the first 10 minutes after the deposition.
\\

\begin{figure}
\includegraphics[width=8cm]{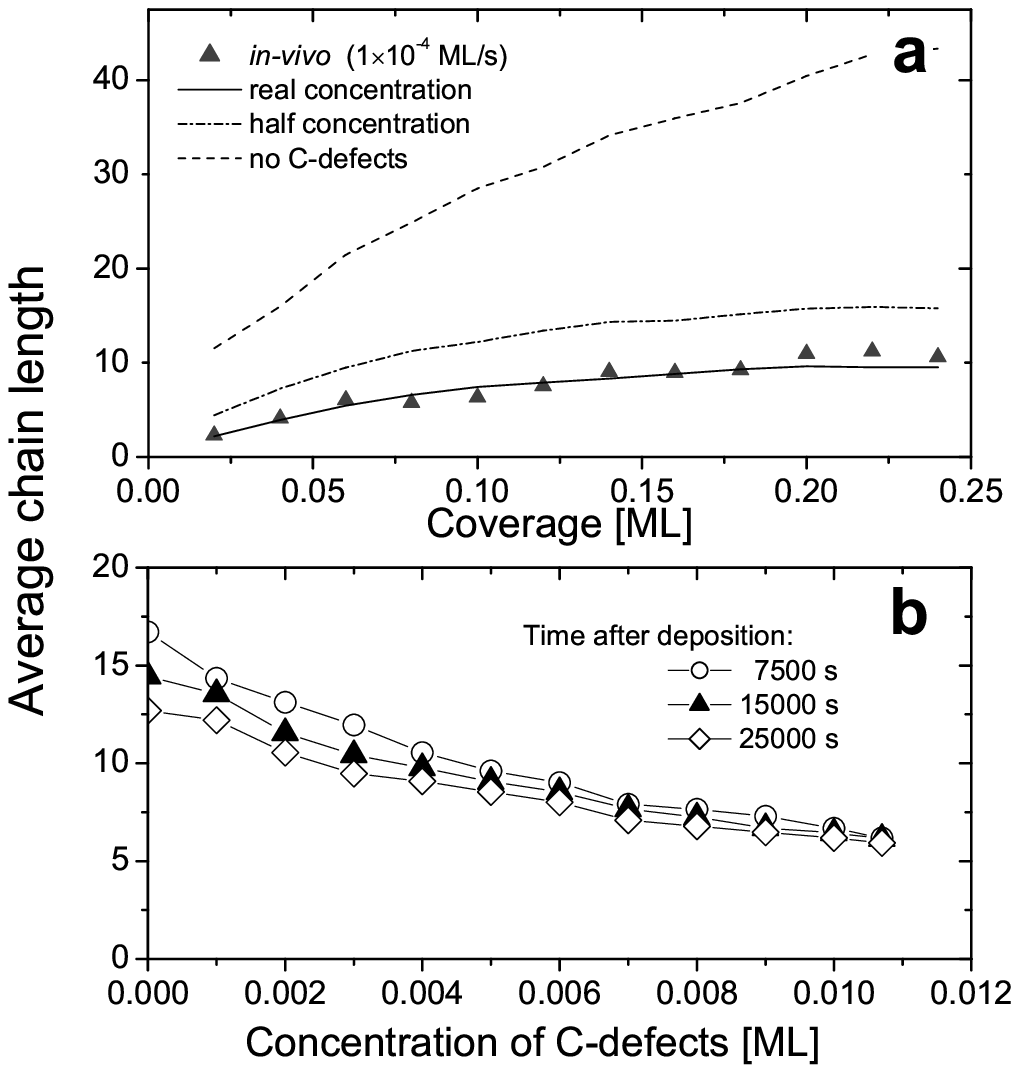}
\caption{a) Dependence of average chain length on coverage simulated for various concentrations of C-defects. Triangles represent data from the corresponding $in$-$vivo$ experiment. b) Average chain length is controlled by concentration of C-defects. Note also the relaxation of average chain length during time. $E_{\|} = 0.62$ eV and $E_{\bot} = 0.58$ eV, $E_{det}= 0.82$ eV.}
\label{C-defects}
\end{figure}

{\it Influence of C-defects}. The time constant for the detachment of an In atom from an adsorption site at a
C-defect is $\approx100\times$ bigger than for the detachment from In chains, making the C-defects practically
perfect diffusion traps \cite{9}. To describe a role of C-defects at heterogeneous nucleation and submonolayer
growth of In quantitatively we simulated growth using the same parameters as for a real experiment, only concentration
of the C-defects was changed. Fig.~\ref{C-defects}a shows dependence of average chain length on coverage obtained
for parameters of the $in~vivo$ experiment with a low deposition rate $1\times10^{-4}$~ML/s. The average chain length
is controlled by concentration of the C-defects. The dependence of the average chain length on the C-defect concentration is on Fig.~\ref{C-defects}b. It can be seen that an effect of postdeposition relaxation disappears with the increasing concentration.
\\

{\it Growth scenario}.  The simulations and analysis of results allow to
suggest a scenario for the experimentally observed submonolayer
growth and discuss a role of participating processes at RT. Simulations
of the postdeposition relaxation show that only a population of atoms
detached from chains exist on the surface after the deposition. They
nucleate as new chains or attach to other chains and the system moves
towards dynamical equilibrium. During the deposition, if the deposition
flux is small enough with respect to detachment rate and surface mobility,
the growth proceeds at thermal equilibrium and a postdeposition effect
consists from thermal fluctuations only -- the scaled chain length
distribution function is monotonously decreasing. At sufficiently high
deposition flux the growth of chains due to attachment of deposited
atoms dominates and the effect of the detachment is suppressed. The
growth becomes irreversible and the scaled distribution function has a
monomodal form (at almost isotropic surface diffusion).\\

{\it Comparison with an equilibrium model}. The model of Tokar and Dryss\'e \cite{13} for equilibrium homogeneous
growth can provide the only parameter, the nearest-neighbor interaction
energy $V^{x}_{NN}$.  It can be determined from the dependence of
average chain length on coverage (using the equation (14) in
Ref.~\onlinecite{13}). In case of the equilibrium model growth
characteristics are independent of a kinetic path. Our data from
$in$-$vivo$ experiments at a rather low deposition rate  may reflect a
situation not too far from an equilibrium state, but the considered growth
is heterogeneous. The data in Fig.~\ref{expscaledfunction}b can be
approximated by an exponential scaling function. The value $V^{x}_{NN}
= - 0.17$~eV obtained from our experimental data for indium is similar
to the pair-interaction energy determined in Ref.~\onlinecite{13} for
gallium ($-0.192$~eV).\\

Simulations of experimental data showed that the deposition rate is a crucial parameter for growing indium on Si(100) surface at RT and determines a transition between irreversible and reversible character of the growth. The measured growth characteristics depend on a process of postdeposition relaxation, which has to be included into the growth simulation.
The monotonous character of the scaled chain length distribution function can be related to a mechanism of atom detachment from the chains. The deposition rate and substrate temperature can be used for controlling competition
between {\it kinetics} and {\it equilibrium}.
The model formulated for simulations is too simplified to be used for explanation all experimentally observed details \cite{Ost_Kinetics_of_Growth} -- for example a plateau in average chain length dependence on coverage within 0.05 and 0.12~ML (see Fig.~\ref{expdata}) obtained from two various $in$-$vivo$ measurements -- but the model explains the most important features of the growth of indium on the surface Si(100)-2$\times$1 at room temperature.

\section{Conclusions}

STM technique was used for studying growth of indium on the
Si(100)-2$\times$1 surface at low coverage and room temperature.
Direct observation during the deposition -- $in$-$vivo$ measurements --
showed that the C-defects act as nucleation centers for indium adatoms.
The majority of indium chain is pinned on one or both ends by a
C-defect,
that determines
the average chain length for a given
coverage. The $in$-$vivo$ observations further revealed the reversible
character of the growth due to atom detachment from the chains.
Statistical characteristics of the In layers (average chain length, average length of chains not terminated
at C-defects, 
dependence of average chain length on coverage, scaled
island-size distribution function) were obtained from the experiments of two types -- the
$in$-$vivo$ measurements and the standard $in$-$situ$ observations
after the deposition of various coverages.

The atomistic model with anisotropic diffusion which includes  presence
of C defects on the surface as well as detachment of atoms from the chain was
developed.
Both $in$-$vivo$ and $in$-$situ$ experiments were simulated
using KMC method.
The simulations showed that the process of
atom detachment can explain the monotonously decreasing shape of the
scaled chain length distribution function.
Free parameters of the model - activation energies for
anisotropic surface diffusion - were determined by comparison of experimental
and simulated characteristic of the indium layers.
The values obtained for the
activation energies (see Table~\ref{tab:table2}) correspond to almost
isotropic surface diffusion in contrary with anisotropic data reported for
the same group metals Ga and Al earlier.

\section*{Acknowledgements}
The work is a part of the research plan MSM 0021620834 that is financed by Ministry of Education of the Czech Republic and was partly supported by projects GACR 202/06/0049, AVOZ 10100520 of ASCR, GAUK 227/2006/B, GAUK 225/2006/B and GAUK 100907. The access to the METACentrum computing facilities provided under the research intent MSM6383917201 is highly appreciated.

\section*{Appendix}
The deviations between experimental and simulated characteristics were determined as follows:
\\
(i) for the average chain length $\langle s \rangle$:
\begin{equation}
D_{i} = (\langle s \rangle _{simulated}- \langle s \rangle _{experiment})^2;
\label{4aberror}
\end{equation}\\  	
(ii) for the dependence of average length on coverage:
\begin{equation}
	D_{ii} = \sum^{\Theta=0.25\;ML}_{\Theta=0.025\;ML}\frac{1}{\sigma^2_\Theta}(\langle s \rangle_{\Theta}^{simulated}- \langle s \rangle s_{\Theta}^{experiment})^2,
	\label{inviverror}
\end{equation}
where $\sigma^2_\Theta$ is a weight parameter -- mean square deviation of chain length at a given coverage $\Theta$ obtained from simulation experiments;\\
\\
(iii) for the scaled chain length distribution functions:
\begin{equation}
	D_{iii} = \sum^{s=\infty}_{s=1}(f^{simulated}_s-f^{experiment}_s)^2,
	\label{scalederror}
\end{equation}
where function values $f_{s}$ are calculated for each chain length of $s$ atoms contained in the data.


\end{document}